\documentclass[sigconf]{acmart}
\usepackage{multirow}   

\copyrightyear{2025}
\acmYear{2025}
\setcopyright{cc}
\setcctype{by}

\acmConference[CIKM '25]{The 34th ACM International Conference on Information and Knowledge Management}{November 10--14, 2025}{Seoul, Republic of Korea}

\acmBooktitle{Proceedings of the 34th ACM International Conference on Information and Knowledge Management (CIKM '25), November 10--14, 2025, Seoul, Republic of Korea}

\acmDOI{10.1145/3746252.3761222}
\acmISBN{979-8-4007-2040-6/2025/11}

\begin{document}

\title{Modality Alignment with Multi-scale Bilateral Attention for Multimodal Recommendation}

\author{Kelin Ren}
\orcid{0009-0000-8230-3299}
\affiliation{%
  \department{Department of Computer Science and Engineering}
  \institution{Hanyang University}
  \city{Ansan}
  \country{Republic of Korea}
}
\email{renkelin@hanyang.ac.kr}

\author{Chan-Yang Ju}
\authornote{Major in Bio Artificial Intelligence.}
\orcid{0000-0002-5064-802X}
\affiliation{%
  \department{Department of Applied Artificial Intelligence}
  \institution{Hanyang University}
  \city{Ansan}
  \country{Republic of Korea}
}
\email{karunogi@hanyang.ac.kr}

\author{Dong-Ho Lee}
\authornote{Corresponding author.}
\orcid{0000-0003-0305-9182}
\affiliation{
  \department{Department of Applied Artificial Intelligence}
  \institution{Hanyang University}
  \city{Ansan}
  \country{Republic of Korea}
}
\email{dhlee72@hanyang.ac.kr}

\renewcommand{\shortauthors}{Kelin Ren, Chan-Yang Ju, and Dong-Ho Lee}

\begin{abstract}
Multimodal recommendation systems are increasingly becoming foundational technologies for e-commerce and content platforms, enabling personalized services by jointly modeling users’ historical behaviors and the multimodal features of items (e.g., visual and textual). However, most existing methods rely on either static fusion strategies or graph-based local interaction modeling, facing two critical limitations: (1) insufficient ability to model fine-grained cross-modal associations, leading to suboptimal fusion quality; and (2) a lack of global distribution-level consistency, causing representational bias. To address these, we propose MambaRec, a novel framework that integrates local feature alignment and global distribution regularization via attention-guided learning. At its core, we introduce the Dilated Refinement Attention Module (DREAM), which uses multi-scale dilated convolutions with channel-wise and spatial attention to align fine-grained semantic patterns between visual and textual modalities. This module captures hierarchical relationships and context-aware associations, improving cross-modal semantic modeling. Additionally, we apply Maximum Mean Discrepancy (MMD) and contrastive loss functions to constrain global modality alignment, enhancing semantic consistency. This dual regularization reduces mode-specific deviations and boosts robustness. To improve scalability, MambaRec employs a dimensionality reduction strategy to lower the computational cost of high-dimensional multimodal features. Extensive experiments on real-world e-commerce datasets show that MambaRec outperforms existing methods in fusion quality, generalization, and efficiency. Our code has been made publicly available at \url{https://github.com/rkl71/MambaRec}.
\end{abstract}




\maketitle

\begin{figure}[!t]
  \centering
  \includegraphics[width=\linewidth]{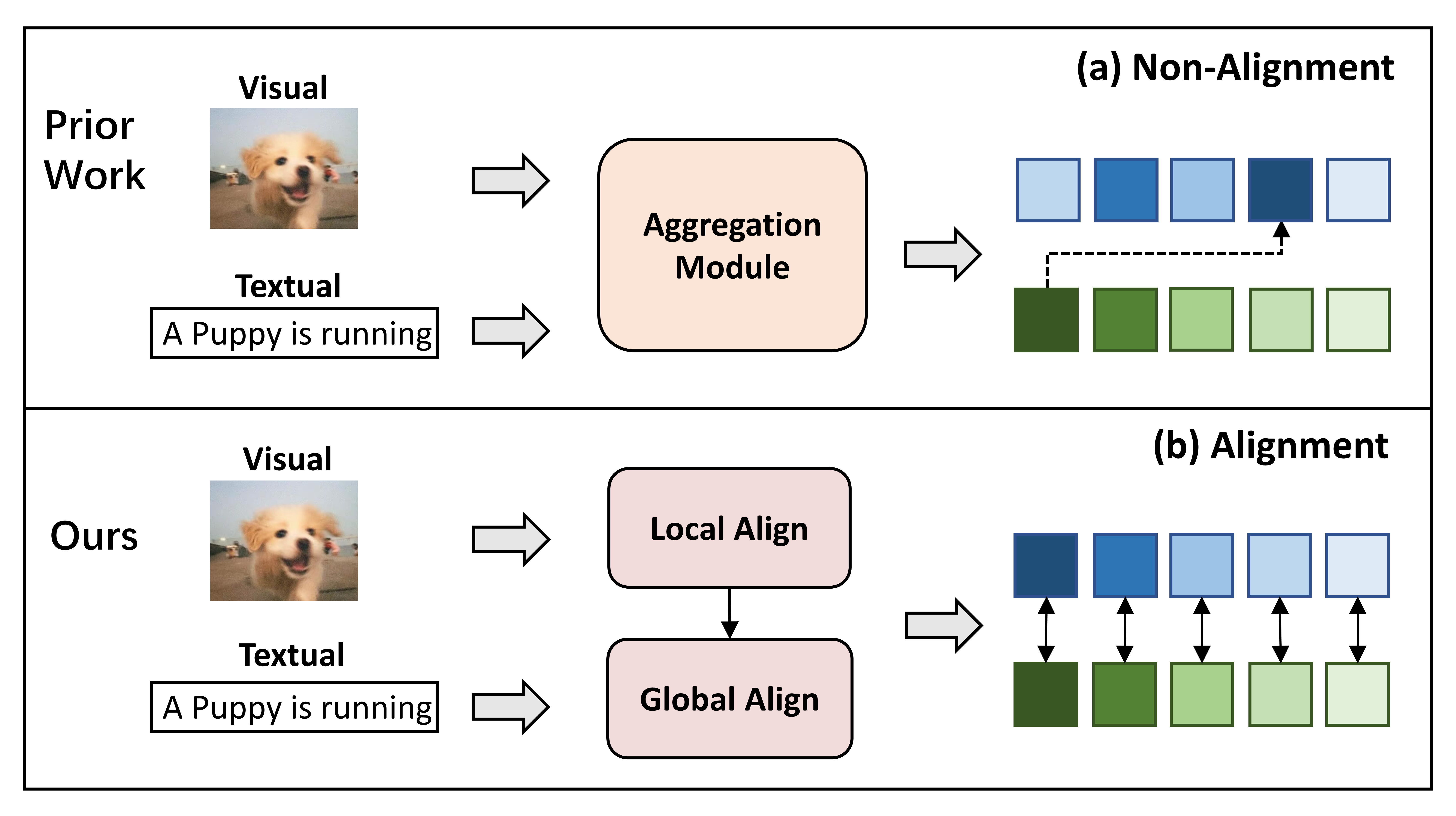}
  \caption{(a) illustrates the process in which existing methods perform simple processing of visual and textual data through aggregation modules, where the resulting feature representations (blue and green blocks) lack alignment. In contrast, (b) introduces Local Alignment and Global Alignment modules to establish a direct, one-to-one correspondence between visual and textual features.}
  \Description{The figure is divided into two parts: the upper part illustrates Prior Work, while the lower part presents Ours. (a) The upper section depicts how conventional methods process visual and textual information. An image of a puppy and the sentence “A puppy is running” are respectively passed through an Aggregation Module, resulting in two sets of feature vectors (represented by blue and green blocks). However, these feature vectors lack explicit correspondence and can only be coarsely aggregated.(b) The lower section illustrates the alignment process proposed in this work. Both the image and text are first processed through a Local Alignment module, followed by a Global Alignment module. The final output consists of two sets of feature vectors, where each visual feature block is explicitly aligned with its corresponding textual feature block via bidirectional arrows.}
\end{figure}

\section{Introduction}
With the rapid development of information technology, recommendation systems have been widely used in various fields such as e-commerce and Short Video platforms, which can effectively help users discover interesting content. Because user decisions are often influenced by the rich multimodal information (such as images and text) presented by items, effective fusion of these multimodal features to accurately capture user preferences has become an important research direction in the recommendation field. Numerous studies have shown that multimodal recommendation systems (MRS) that integrate modal information generally perform better than traditional recommendation methods that rely solely on historical interaction data \cite{he2016vbpr, zhang2021mining, 10.1145/3695461, zhou2023comprehensive, kim2024monet}. In particular, multimodal information can supplement sparse or missing behavioral data by providing semantically rich content from auxiliary sources, thereby alleviating the cold start and data sparse issues inherent in many real-world recommendation scenarios. The increasing popularity of user-generated content and high-dimensional multimedia data makes it not only desirable but also necessary to efficiently and effectively model cross-modal relationships for next-generation recommendation systems.

In recent years, multimodal recommendation methods based on Graph Convolutional Networks (GCNs) have achieved good results, mainly because GCNs can effectively capture complex interactive information and semantic relationships between users and items \cite{wei2019mmgcn, wei2020graph, liu2023multimodal}. The classic Multi-View Graph Convolutional Network (MGCN) \cite{yu2023multi} builds a collaboration graph between users and items through feature mapping of different modes, thereby better reflecting the relationship between users and items \cite{yu2023multi}. Despite notable progress, existing multimodal recommendation methods still suffer from several limitations. In the feature extraction stage, many models rely on static linear projections or simplistic fusion techniques, which limit their ability to capture fine-grained semantic correspondence between visual and textual modalities. Models like Bootstrap Multimodal Matching (BM3) \cite{zhou2023bootstrap} and Self-supervised Learning for Multimedia Recommendation (SLMRec) \cite{tao2022self} introduce contrastive objectives or local fusion, yet typically lack global constraints on modality-level representation distributions, leading to semantic misalignment. Moreover, graph-based approaches such as Local and Global Graph Learning for Multimodal Recommendation (LGMRec) \cite{guo2024lgmrec} utilize modality-specific graphs to model preferences, but still struggle with the distribution shift across modalities \cite{cai2024multimodal} and the noise introduced by high-dimensional features \cite{yang2023bicro}. These limitations hinder the effectiveness and generalization of multimodal recommendation systems in large-scale scenarios.

To address these challenges, we propose a \textbf{M}odality \textbf{A}lignment with \textbf{M}ulti-scale \textbf{B}ilateral \textbf{A}ttention (MAMBARec) for multimodal recommendations. The model is designed to achieve more efficient and accurate feature fusion and alignment. Specifically, this study introduces three key components to enhance multimodal feature alignment:

In order to improve the accuracy of local feature alignment, we propose a \textbf{D}ilated \textbf{RE}finement \textbf{A}ttention \textbf{M}odule (DREAM), which uses multi-scale dilated convolution to capture modal feature information at different scales, and combines channel and spatial attention mechanisms to adaptively highlight key features. The DREAM module can effectively expand the receptive field without increasing the computational burden, allowing the model to focus on fine-grained local interactions and broader context dependencies at the same time, which is critical to modeling fine preferences in heterogeneous modal spaces.

Moreover, in order to effectively enforce the consistency of the global distribution of modal characteristics, we introduce the Maximum Mean Discrepancy (MMD) loss function \cite{gretton2012kernel} into the recommendation area. By using the Gaussian kernel function to explicitly restrict the consistency of modal feature distribution, the quality of feature fusion is significantly improved. Finally, in order to solve the memory consumption problem caused by high-dimensional characteristics, we designed a novel memory optimization strategy. By introducing configurable dimension reduction factors and a two-stage feature conversion mechanism, this strategy significantly reduces memory usage while maintaining high-quality feature representations. This ensures that MAMBARec remains scalable and computationally efficient, allowing it to be deployed in large-scale recommendation environments without sacrificing performance. Comprehensive experiments on three public datasets demonstrate the significant advantages of our proposed method.

Our main contributions can be summarized as follows:
\begin{itemize}
    \item We propose an innovative DREAM module that significantly enhances local feature alignment through multi-scale extended convolution and dual attention mechanisms.
    \item We also introduce the MMD loss function into the recommendation system to reduce modal discrepancies and enhance the model’s generalization, by enforcing consistency in the global distribution of modal features.
    \item Furthermore, we designed and implemented a flexible and efficient memory optimization strategy to effectively reduce the memory requirements of high-dimensional modal features and improve the actual deployment capabilities of the model.
\end{itemize}

\section{Related Work}
Traditional recommendation systems primarily relied on collaborative filtering methods to learn latent user preferences from user-item interaction data. For instance, the Bayesian Personalized Ranking (BPR) model \cite{rendle2012bpr} optimizes item ranking using pairwise ranking loss functions, enhancing recommendation accuracy. However, collaborative filtering faces challenges such as data sparsity and cold-start problems, limiting its ability to capture diverse user interests. To address these, some approaches incorporate multimodal features, providing richer contextual information. The Visual Bayesian Personalized Ranking (VBPR) model \cite{he2016vbpr} integrates visual features extracted from pre-trained Convolutional Neural Networks (CNNs), mitigating cold-start issues and enhancing recommendation performance. However, early multimodal approaches often treat each modality independently, using simple linear combinations that fail to model complex cross-modal interactions, limiting their ability to capture fine-grained semantic alignments.

\begin{figure*}[t]
  \centering
  \includegraphics[width=\textwidth]{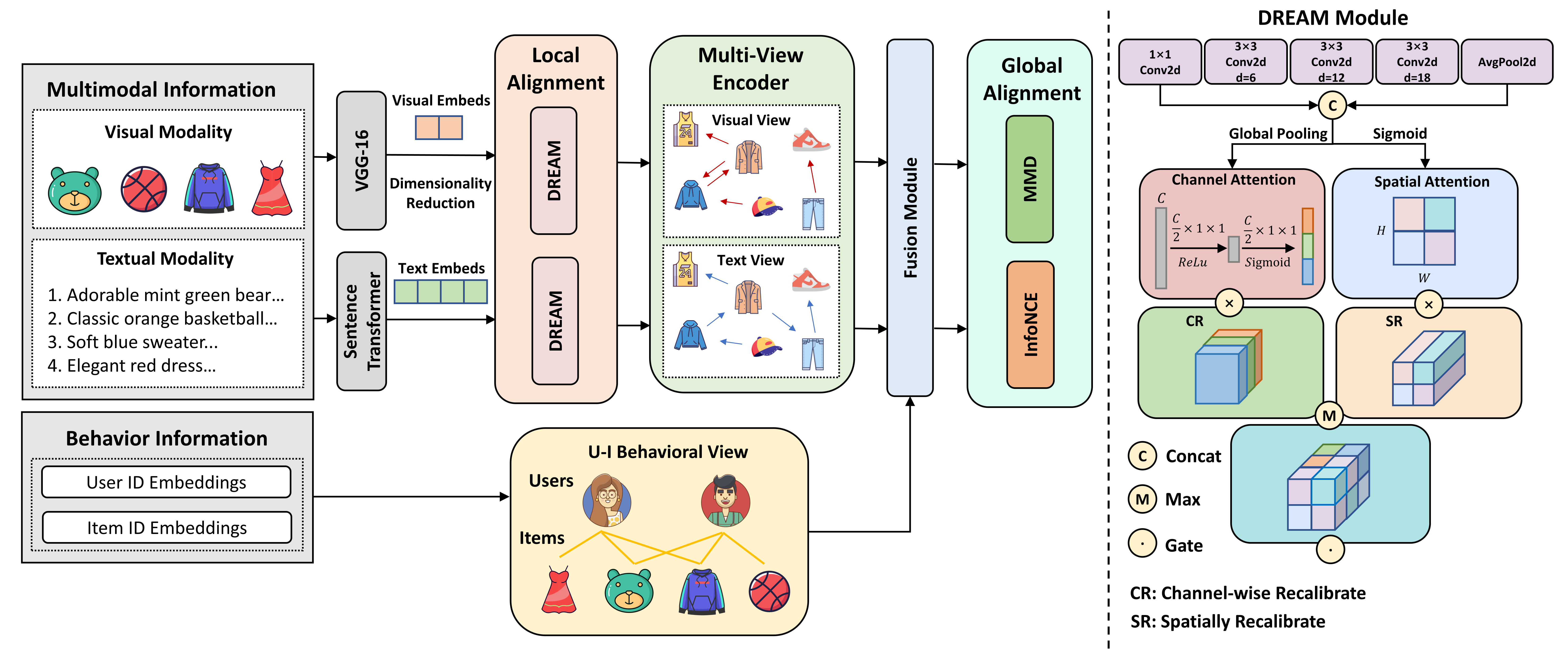}
  \caption{The overall architecture of the proposed MambaRec, comprising three key components. (i) The local alignment performs fine-grained matching between local features of images and text. (ii) The global alignment ensures consistency between different modalities at the level of global distribution. (iii) The dimensionality reduction mechanism was introduced before alignment to compress high-dimensional visual features and reduce memory overhead. }
  \Description{The figure is organized into three main sections from left to right, with the detailed structure of the DREAM module highlighted on the far right.}
  \label{fig:teaser}
\end{figure*}

The advent of Graph Neural Networks (GNNs) has presented new opportunities for multimodal recommendation systems. LightGCN \cite{he2020lightgcn} introduces a simplified graph convolution framework focused on neighborhood aggregation, which efficiently models high-order user-item relationships. The Multimodal Graph Convolutional Network (MMGCN) \cite{wei2019mmgcn} further advances this by constructing interaction graphs for each modality and fusing their graph embeddings. While this fusion improves recommendation performance, it may amplify modality-specific noise. GRCN \cite{wei2020graph} addresses this by refining the graph through gating mechanisms to suppress noisy interactions. However, existing GCN-based models still struggle with accommodating heterogeneous semantic granularities across modalities, leading to information dilution and over-smoothing in dense graphs \cite{li2018deeper}.

Recent work in multimodal recommendation has further explored self-supervised learning and fine-grained fusion mechanisms. SLMRec \cite{tao2022self} introduces cross-modal self-supervised tasks and a hierarchical contrastive learning framework to enhance the robustness of modal representations. BM3 \cite{zhou2023bootstrap} proposes a self-supervised strategy that generates contrastive views through random feature dropping, reducing reliance on negative sampling while improving both efficiency and accuracy. FREEDOM \cite{zhou2023tale} reduces memory consumption by freezing the item similarity graph and pruning user interaction graphs, aiming to improve training stability and resource efficiency. LGMRec \cite{guo2024lgmrec} introduces a local–global modal graph learning strategy that models both modality-specific user interests and shared cross-modal preferences, in order to alleviate interference between modalities. Meanwhile, MGCN \cite{yu2023multi} leverages users’ historical behavior to guide the preprocessing and refinement of modal features and employs multi-view graph convolution to facilitate more fine-grained fusion of multimodal information. While these approaches represent important progress in multimodal recommendation, they often overlook semantic consistency across modalities and coordination of representation distributions. Few approaches explicitly integrate local feature alignment and distributional consistency in a unified manner, which is crucial for achieving robust and generalized cross-modal fusion. In this study, we propose to jointly model local semantic correspondences and global distributional consistency from a multimodal alignment perspective to achieve more consistent and efficient recommendation systems.

\section{Problem Formulation}
Given a set of $M$ users and $N$ items, the historical interactions between users and items are represented by a sparse binary matrix $\mathbf{R} \in \{0,1\}^{M \times N}$, where $\mathbf{R}_{ui} = 1$ indicates that user $u$ has interacted with item $i$ (e.g., by clicking or purchasing), and $\mathbf{R}_{ui} = 0$ otherwise. To capture collaborative signals, each user and item is assigned a $d$-dimensional identity embedding, denoted as $\mathbf{p}_u$ and $\mathbf{q}_i$, respectively. In addition, each item is associated with multimodal features. This work focuses on two modalities, namely the visual mode ($v$) and the text mode ($t$), where the set of modalities is defined as $\mathcal{M} = \{v, t\}$. For each modality $m \in \mathcal{M}$, the raw feature of item $i$ is represented as $\mathbf{x}_i^m \in \mathbb{R}^{d_m}$ and is projected into a unified embedding space using a learnable linear transformation $W_m \in \mathbb{R}^{d \times d_m}$, resulting in the modality embedding $\mathbf{h}_i^m = W_m \mathbf{x}_i^m$. The predicted preference score between user $u$ and item $i$ is computed using a scoring function $\hat{y}_{ui} = f(\mathbf{p}_u, \mathbf{q}_i, \{\mathbf{h}_i^m\}_{m \in \mathcal{M}})$, where $f$ can be a simple inner product or a deep model designed to fuse collaborative and multimodal information. The model is trained using observed interactions as supervision, and during inference, each user receives a personalized recommendation list based on the predicted scores. Although this paper focuses on both visual and textual modalities, the proposed method is well generalizable. We aim to accurately predict the interaction probability $\hat{y}_{ui}$ between the user $u$ and the item $i$ for personalized recommendations.

\section{Methodology}
The key components of MamabaRec include three core aspects: (i) Local Feature Alignement, (ii) Global Distribution Alignment, and (iii) Dimensionality Optimization Mechanism. Figure. 2 shows an overview of the proposed architecture.

\subsection{Local Feature Alignment}
Prior work \cite{li2021align} has highlighted the importance of fusing semantically aligned image and text features. Inspired by DeepLabV3 with Atrous Spatial Pyramid Pooling (ASPP) \cite{chen2017rethinking} and the Convolutional Block Attention Module (CBAM) \cite{woo2018cbam}, we design a Dilated REfinement Attention Module (DREAM) to achieve local feature alignment through multi-scale convolution and attention mechanisms.
Let the input feature map be $X \in \mathbb{R}^{C \times H \times W}$ and the output be $Y \in \mathbb{R}^{C \times H \times W}$. The DREAM module contains five parallel convolution operation branches to extract multi-scale features, which are then weighted and fused by simultaneously applying channel and spatial attention mechanisms to the fused feature map.

\subsubsection{Multi-scale feature extraction}
These five branches consist of a $1 \times 1$ convolution branch, three dilated convolution branches with increasing dilation rates, and a global average pooling branch. The first branch applies a $1 \times 1$ convolution to extract fine-grained local features with minimal computational cost and without expanding the receptive field. This enables lightweight channel-wise transformation while preserving high-resolution spatial details. Branches 2 to 4 employ $3 \times 3$ dilated convolutions with dilation rates of 6, 12, and 18, respectively. These configurations enlarge the receptive field to capture medium- and long-range contextual information while maintaining the spatial resolution through appropriate padding. Branch 5 performs global average pooling over the entire spatial dimension, followed by a $1 \times 1$ convolution to project the global representation into the desired number of channels. The resulting $1 \times 1$ feature map is then upsampled to $H \times W$ via bilinear interpolation to align with the spatial dimensions of the other branches and provide global semantic context. Let the output feature maps of the five branches be denoted as $F_1, F_2, F_3, F_4 \text{ and } F_5 \in \mathbb{R}^{H \times W \times C_i}$. There are concatenated along the channel dimension to obtain the fused feature map:
\begin{equation}
F = \operatorname{Concat}_c(F_1, F_2, F_3, F_4, F_5),
\end{equation}
where $\operatorname{Concat}_c$ denotes concatenation along the channel dimension. The resulting fused representation $F \in \mathbb{R}^{H \times W \times C}$ integrates both local detail and global context information, where $C = \sum_{i=1}^{5} C_i$.

\subsubsection{Channel attention mechanism}
In order to adaptively enhance important channels, DREAM introduced a channel attention module to the fused feature map $F$. Apply global average pooling to the feature map $F$ to obtain a channel compression vector $z \in \mathbb{R}^{C'\times 1 \times 1}$ (if there are $C'$channels after splicing). Then, enter $z$ into the two fully connected layers in turn. The first fully connected layer reduces the channel dimension (using the ReLU activation function $\delta(\cdot)$ to $C'/ r$), and the second fully connected layer restores the dimension to $C'$ and outputs a channel weight vector $M_c \in \mathbb{R}^{C'\times 1 \times 1}$. The above process can be expressed as:
\begin{equation}
M_c = \sigma\left(W_2\left(\delta\left(W_1(z)\right)\right)\right),
\end{equation}
where $W_1$ and $W_2$ are the weight matrices of the fully connected layer, and $\delta$, $\sigma$ denote the ReLU and Sigmoid activation functions respectively. The value range of each element of the obtained channel attention weight $M_c$ is $[0,1]$, which indicates the importance of the corresponding channel. Finally, the weight $M_c$is multiplied back onto the fused feature map to achieve channel-by-channel feature recalibration:
\begin{equation}
F_c = F \otimes M_c,
\end{equation}
where the symbol $\otimes$ denotes element-wise multiplication for each channel. After channel attention is enhanced, the feature map $F_c$ is enlarged on key channels to suppress unimportant channel information.

\subsubsection{Spatial attention mechanism}
The DREAM module applies spatial attention mechanisms to highlight key spatial locations. Specifically, the fused feature $F$ is pooled in the channel dimension to obtain a two-dimensional feature map $P \in \mathbb{R}^{1 \times H \times W}$. Here we use an average per channel method, $P(i, j) = \frac{1}{C'}\sum_{c=1}^{C'} F(c, i, j)$, and calculate the average response at each spatial location $(i, j)$. Next, $P$ is linearly transformed using a convolution of $1\times1$, and then normalized through a Sigmoid function to get a spatial attention weight graph $M_s \in \mathbb{R}^{1 \times H \times W}$:
\begin{equation}
M_s = \sigma\left(f^{1 \times 1}(P)\right),
\end{equation}
where $f^{1\times1}$ denotes the $1\times1$ convolution operator, and $\sigma$ maps the result to the $[0,1]$ interval. Each position $(i,j)$ of $M_s$ corresponds to a spatial weight, which indicates the importance of the position. This weight map is then expanded to match the number of channels in $F$ (copied to all channels at each position) and multiplied element-wise with the original fused feature map:
\begin{equation}
F_s = F \otimes M_s,
\end{equation}
this produces a feature map $F_s$with enhanced spatial attention. After this operation, $F_s$ is enhanced at important spatial locations, suppressing the influence of background or irrelevant areas.

\subsubsection{Attention fusion}
After obtaining the channel enhancement feature $F_c$ and the spatial enhancement feature $F_s$, the DREAM module fuses the two features by performing an element-by-element maximization operation, thereby making full use of these two types of attention information. Specifically, we perform an element-by-element maximum operation on $F_c$ and $F_s$(denoted as $\mathcal{M}$) to merge them:
\begin{equation}
Y = \mathcal{M}(F_c,\; F_s),
\end{equation}
where $\mathcal{M}(a, b)$ means selecting the larger of the two input tensors $a$ and $b$ element by element. For each position in each channel, the more responsive one of the channel attention branches or the spatial attention branches is adaptively selected as the output. After the DREAM module performs multi-scale feature refinement and attention-weighted alignment, more significant and accurate local feature information is included.

\subsection{Global Attribution Alignment}
The global alignment module is used to achieve consistency in the spatial representation distribution of image and text features, with the Maximum Mean Discrepancy (MMD) loss \cite{gretton2012kernel} as the core, supplemented by InfoNCE comparison loss \cite{oord2018representation} for further feature alignment. Specifically, assuming that the image modal feature set is $V = \{v_i\}_{i=1}^{N}$ and the text modal feature set is $T = \{t_j\}_{j=1}^{N}$, we use MMD to measure the difference between the two modal feature distributions. MMD uses the embedding gap of the distribution mean in the reproducing kernel Hilbert space (RKHS) to measure the inconsistency of the distribution. It is defined as follows,
\begin{equation}
\mathrm{MMD}^2(V, T) = \left\| \frac{1}{N} \sum_{i=1}^{N} \phi(v_i) - \frac{1}{N} \sum_{j=1}^{N} \phi(t_j) \right\|_{\mathcal{H}}^2,
\end{equation}
where $\phi(\cdot)$denotes the mapping function in the reproducing kernel Hilbert space (RKHS). This expression can be equivalently converted to a kernel function form:
\begin{align}
\mathrm{MMD}^2(V, T) &= \frac{1}{N^2} \sum_{i=1}^{N} \sum_{i'=1}^{N} k(v_i, v_{i'}) 
+ \frac{1}{N^2} \sum_{j=1}^{N} \sum_{j'=1}^{N} k(t_j, t_{j'}) \notag \\
&\quad - \frac{2}{N^2} \sum_{i=1}^{N} \sum_{j=1}^{N} k(v_i, t_j)
\end{align}
where $k(\cdot,\cdot)$ is the kernel function defined on the input space. The Gaussian kernel function we use in this work is defined as follows:
\begin{equation}
k(v, t) = \exp\left(-\frac{\left\| v - t \right\|^2}{2\sigma^2}\right),
\end{equation}
where $\sigma$ is a kernel bandwidth hyperparameter, which is used to adjust the sensitivity of distance between features to similarity. By minimizing the above-mentioned MMD loss, the distance between image modality and text modality in the overall distribution can be effectively shortened, thereby achieving global semantic alignment.

To further enhance the consistency of representations, we introduced InfoNCE comparative learning loss as a supplement. It enhances the similarity between matching pairs of graphic features and suppresses the similarity between non-matching pairs through comparative learning, which is expressed as follows:
\begin{equation}
\mathcal{L}_{\text{InfoNCE}} = -\frac{1}{N} \sum_{i=1}^{N} \log \frac{\exp\left(\mathrm{sim}(v_i, t_i)/\tau\right)}{\sum_{j=1}^{N} \exp\left(\mathrm{sim}(v_i, t_j)/\tau\right)},
\end{equation}
where $\text{sim}(\cdot,\cdot)$ is the dot product of the normalized feature vector, and $\tau$is the temperature hyperparameter.
Ultimately, the overall loss of the global distribution alignment module consists of a weighted combination of the two components:
\begin{equation}
\mathcal{L}_{\text{align}} = \lambda_{\text{mmd}} \cdot \mathcal{L}_{\text{MMD}} + \lambda_{\text{cl}} \cdot \mathcal{L}_{\text{InfoNCE}},
\end{equation}
where $\lambda_{\text{mmd}}$ and $\lambda_{\text{cl}}$ denote the weight coefficients of these two losses, and their values are specified by the model configuration hyperparameter, respectively. With this joint optimization strategy, we achieved alignment and fusion of cross-modal semantic representations at the global distribution level.

\subsection{Dimensionality Optimization}
In order to effectively solve the dimensional differences and redundancy problems between different modal (image and text) features, we propose a dimensional optimization mechanism to improve the efficiency of multimodal feature fusion and the generalization ability of the model. Let the image modal feature matrix be $V \in \mathbb{R}^{N \times D_V}$, and the text modal feature matrix be $T \in \mathbb{R}^{N \times D_T}$, where $N$ denotes the number of samples, and $D_V$, $D_T$ denote the original feature dimensions of the image and text respectively. We introduce two linear dimensionality reduction matrices $W_V \in \mathbb{R}^{D_V \times d}$ and $W_T \in \mathbb{R}^{D_T \times d}$, which are used to project the two modal features respectively:
\begin{equation}
V' = V W_V, \quad T' = T W_T,
\end{equation}
where $V'$,$T'\in \mathbb{R}^{N \times d}$ denote the unified feature representation after dimension reduction, and $d$ is the target embedding dimension, whose size is controlled by the dimension reduction coefficient $r$, which is defined as follows:
\begin{equation}
d = \left\lfloor \frac{\min(D_V, D_T)}{r} \right\rfloor.
\end{equation}
The dimensionality reduced modal features $V'$ and $T'$ will be further input into the DREAM module for structural enhancement and feature fine-grained enhancement. Overall, this dimension optimization mechanism can not only significantly reduce the dimension and computing overhead of the feature space, but also improve the robustness of the multimodal fusion representation.

\subsection{Multi-View Encoder}
To enhance the representation of specific modes before global alignment, we adopt a multi-view encoder architecture inspired by previous work MGCN \cite{yu2023multi}. This encoder processes the output of the Local Alignment Module (DREAM) to generate two semantically rich views, a visual view and a text view. Each view captures intra-modal relationships and complementary semantics that may be lost during local fusion.

\subsection{Prediction and Optimization}
In this model, for any user $u$ and item $i$, the model generates their fusion representation vectors $e_u^* \in \mathbb{R}^d$ and $e_i^* \in \mathbb{R}^d$ respectively. The interaction scores of the two are calculated through the inner product of the vectors:
\begin{equation}
\hat{y}(u,i) = (e_u^*)^\top e_i^*,
\end{equation}
Model training uses Bayesian Personalized Ranking (BPR) \cite{rendle2012bpr} as the basic optimization goal. This goal is used to characterize the user's preference ranking relationship for positive and negative sample items. The loss function is defined as follows:
\begin{equation}
\mathcal{L}_{\mathrm{BPR}} = \sum_{(u,i,j)\in \mathcal{O}} -\log\sigma(\hat{y}_{ui} - \hat{y}_{uj}),
\end{equation}
where $\sigma(\cdot)$ denotes the Sigmoid function, and $\mathcal{O} = \{(u, i, j)\}$ denotes the set of triples where user $u$ prefers positive samples $i$ over negative samples $j$.

To enhance the consistency of representation learning and cross-modal fusion, the model further introduces InfoNCE comparative learning loss $\mathcal{L}_{\mathrm{cl}}$ and maximum mean discrepancy loss $\mathcal{L}_{\mathrm{mmd}}$ as auxiliary optimization terms. Introduce $\ell_2$ regularization to prevent overfitting the model. The final joint optimization goals are as follows:
\begin{equation}
\mathcal{L} = \mathcal{L}_{\mathrm{BPR}} 
+ \lambda_{\mathrm{cl}} \cdot \mathcal{L}_{\mathrm{cl}} 
+ \lambda_{\mathrm{mmd}} \cdot \mathcal{L}_{\mathrm{mmd}} 
+ \lambda_{\mathrm{reg}} \cdot \|\Theta\|_2^2,
\end{equation}
where $\Theta$ denotes the set of all learnable parameters in the model, and $\lambda_{\mathrm{cl}}$, $\lambda_{\mathrm{mmd}}$, $\lambda_{\mathrm{reg}}$ are weight hyperparameters of the corresponding loss term.

\section{Experiments}
We conduct a series of experiments to address the following research questions:

\begin{itemize}
\item {\texttt{\textbf{RQ1:}}} How effective is the proposed MambaRec architecture compared to state-of-the-art general-purpose and multimodal recommendation models?
\item {\texttt{\textbf{RQ2:}}} What are the contributions of key components and individual modalities within MambaRec to its overall performance?
\item {\texttt{\textbf{RQ3:}}} How does the variation of hyperparameters affect the overall effectiveness of the proposed model?
\item {\texttt{\textbf{RQ4:}}} Why does modality alignment lead to improved recommendation performance?
\end{itemize}

\begin{table}[h] 
  \caption{Statistics of the experimental datasets}
  \label{tab:datasets}
  \centering
  \begin{tabular}{lrrcr}
    \toprule
    Dataset & \#User & \#Item & \#Interaction & Density \\
    \midrule
    Baby & 19,445 & 7,050 & 160,792 & 0.117\% \\
    Sports & 35,598 & 18,357 & 296,337 & 0.045\% \\
    Clothing & 39,387 & 23,033 & 278,677 & 0.031\% \\
    \bottomrule
  \end{tabular}
\end{table}

\subsection{Experimental Settings}
\subsubsection{Datasets} 
Following \cite{zhou2023tale, guo2024lgmrec}, We use three representative subsets of Amazon's product review datasets \footnote{Datasets are available at \url{http://jmcauley.ucsd.edu/data/amazon/links.html}} for experimental evaluations: (a) Baby, (b) Sports and Outdoors, and (c) Clothing, Shoes and Jewelry. For ease of reference, we simply refer to these three categories as Baby, Sports and Clothing. These three categories cover different consumption scenarios, have significant user behavior differences and pattern characteristics diversity. The raw data are filtered based on the 5-core setting on both users and items, and previous works mostly use these processed subsets for multimodal recommendation. The processing results are shown in Table 1. For the visual modality, we first use 4,096-dimensional visual features extracted by a pre-trained VGG16 convolutional neural network \cite{simonyan2014very} followed by dimensionality reduction. For the textual modality, we extract a 384-dimensional textual embedding by utilizing sentence-transformers \cite{reimers2019sentence} on the concatenation of the title, descriptions, categories, and brand of each item.

\begin{table*}[h]
  \centering
  \caption{The overall performance of MambaRec and other baseline models on three datasets. The best result shown in boldface and the next-best is underlined. The t-tests verified the significance of performance improvements with $p$-value $<$ 0.01.}
  \label{tab:performance}
  \setlength{\tabcolsep}{2pt}      
  \footnotesize                    
  \renewcommand{\arraystretch}{1.2} 

  \resizebox{\textwidth}{!}{%
    \begin{tabular}{llcccccccccc}
      \toprule
      \multirow{2}{*}{Datasets} & \multirow{2}{*}{Metrics}
        & \multicolumn{2}{c}{General Models}
        & \multicolumn{8}{c}{Multimodal Models} \\
      \cmidrule(lr){3-4}\cmidrule(lr){5-12}
      & 
        & BPR & LightGCN
        & VBPR & MMGCN & GRCN & SLMRec & FREEDOM & MGCN & LGMRec & MAMBARec \\
      \midrule
      \multirow{4}{*}{Baby} 
      & Recall@10 & 0.0382 & 0.0453 & 0.0425 & 0.0424 & 0.0534 & 0.0545 & 0.0627 & 0.0620 & \underline{0.0644} & \textbf{0.0660}$^*$ \\
      & Recall@20 & 0.0595 & 0.0728 & 0.0663 & 0.0668 & 0.0831 & 0.0837 & 0.0992 & 0.0964 & \underline{0.1002} & \textbf{0.1013}$^*$ \\
      & NDCG@10   & 0.0214 & 0.0246 & 0.0223 & 0.0223 & 0.0288 & 0.0296 & 0.0330 & 0.0339 & \underline{0.0349} & \textbf{0.0363}$^*$ \\
      & NDCG@20   & 0.0263 & 0.0317 & 0.0284 & 0.0286 & 0.0365 & 0.0371 & 0.0424 & 0.0427 & \underline{0.0440} & \textbf{0.0454}$^*$ \\
      \midrule
      \multirow{4}{*}{Sports} 
      & Recall@10 & 0.0417 & 0.0542 & 0.0561 & 0.0386 & 0.0607 & 0.0676 & 0.0717 & \underline{0.0729} & 0.0720 & \textbf{0.0763}$^*$ \\
      & Recall@20 & 0.0633 & 0.0837 & 0.0857 & 0.0627 & 0.0922 & 0.1017 & 0.1089 & \underline{0.1106} & 0.1068 & \textbf{0.1147}$^*$ \\
      & NDCG@10   & 0.0232 & 0.0300 & 0.0307 & 0.0204 & 0.0325 & 0.0374 & 0.0385 & \underline{0.0397} & 0.0390 & \textbf{0.0416}$^*$ \\
      & NDCG@20   & 0.0288 & 0.0376 & 0.0384 & 0.0266 & 0.0406 & 0.0462 & 0.0481 & \underline{0.0496} & 0.0480 & \textbf{0.0514}$^*$ \\
      \midrule
      \multirow{4}{*}{Clothing} 
      & Recall@10 & 0.0200 & 0.0338 & 0.0281 & 0.0224 & 0.0428 & 0.0461 & 0.0629 & \underline{0.0641} & 0.0555 & \textbf{0.0673}$^*$ \\
      & Recall@20 & 0.0295 & 0.0517 & 0.0410 & 0.0362 & 0.0663 & 0.0696 & 0.0941 & \underline{0.0945} & 0.0828 & \textbf{0.0996}$^*$ \\
      & NDCG@10   & 0.0111 & 0.0185 & 0.0157 & 0.0118 & 0.0227 & 0.0249 & 0.0341 & \underline{0.0347} & 0.0302 & \textbf{0.0367}$^*$ \\
      & NDCG@20   & 0.0135 & 0.0230 & 0.0190 & 0.0153 & 0.0287 & 0.0308 & 0.0420 & \underline{0.0428} & 0.0371 & \textbf{0.0449}$^*$ \\
      \bottomrule
    \end{tabular}%
  } 
  \vspace{4pt}
\end{table*}

\subsubsection{Baselines} 
To evaluate the effectiveness of MambaRec model, we compare it with several state-of-the-art (SOTA) recommendation methods in two categories: 

\noindent \textbf{i)  General Models:}
\begin{itemize}
\item {\texttt{\textbf{BPR} \cite{rendle2012bpr}:}} A classic sorting optimization model based on matrix decomposition.
\item {\texttt{\textbf{LightGCN} \cite{he2020lightgcn}:}} Simplifying the graph convolution structure to efficiently model high-order neighbor relationships. 
\end{itemize}

\noindent \textbf{ii) Multimodal Models:}
\begin{itemize}
\item {\texttt{\textbf{MMGCN} \cite{wei2019mmgcn}, \textbf{GRCN} \cite{wei2020graph}:}} Optimize information dissemination through specific modal graph and graph reconstruction respectively.
\item {\texttt{\textbf{MGCN}  \cite{yu2023multi}:}} Build multimodal collaboration graphs through feature fusion for recommendation.
\item {\texttt{\textbf{SLMRec}  \cite{tao2022self}, \textbf{FREEDOM} \cite{zhou2023tale}, \textbf{LGMRec} \cite{guo2024lgmrec}:}} Capturing cross-modal semantic relationships based on self-supervised or comparative learning.
\end{itemize}

\subsubsection{Evaluation Protocols} 
For a fair comparison, we follow the same evaluation setup as in \cite{zhou2023tale, yu2023multi, guo2024lgmrec}. Specifically, we randomly split each user's interaction history into training, validation, and test sets with a ratio of 8:1:1 to ensure the robustness of the training process and the fairness of the evaluation. During the evaluation phase, we adopt two widely used metrics, Recall@K and NDCG@K, to assess the model's ability to retrieve relevant items and its ranking quality in top-K recommendations. To ensure comparability across different models, we report the average performance across all users in the test set for K = 10 and K = 20.

\subsubsection{Implementation Details} 
We implemented the proposed model and all comparison baseline methods based on the unified open-source framework MMRec \cite{zhou2023mmrec} and evaluated on a GeForce RTX4090 GPU card with 24 GB memory. To ensure the repeatability and fairness of experiments, all models were trained using the Adam optimizer \cite{kingma2014adam} and adjusted with reference to the optimal hyperparameter configurations reported in each baseline paper. In addition, we uniformly set the embedding dimension of users and items to 64, and use the Xavier \cite{glorot2010understanding} initialization method to initialize model parameters. During the training phase, we set the batch size to 2048, the maximum number of training rounds to 1000, and triggered the early stop mechanism when Recall@20 on the validation set failed to improve for 20 consecutive times to avoid overfitting. The learning rate was set to 0.001, with a decay factor of 0.96 every 50 epochs. For regularization, we used a weight decay of \(1 \times 10^{-4}\) and a contrastive learning loss (\(cl\_loss\)) of 0.01. Additionally, we incorporated Maximum Mean Discrepancy (MMD) \cite{gretton2012kernel} with varying weights \([0.1, 0.15, 0.2]\) and kernel bandwidths \([1.0, 1.5, 2.0]\). A reduction factor of 8 was applied during feature extraction to improve model efficiency. To further ensure the stability of the results, the same random seed was used in all implementations of the baseline model.

\subsection{Overall Performance (RQ1)}
Table. 2 shows the performance of the proposed MambaRec model compared with other baseline models on three datasets. From the table, we can clearly see the following results:

(1) \textbf{MambaRec performs significantly better than general recommendation models and multimodal recommendation models.}
Specifically, the core advantage of our model lies in its multi-level modal alignment and fusion mechanism. At the local level, the model introduces a dilated refinement attention module (DREAM), which enhances fine-grained representations of image and text modalities. At the global level, the model explicitly aligns image and text distributions using the Maximum Mean Discrepancy (MMD) constraint, which effectively alleviates cross-modal semantic bias. In addition, through channel attention and spatial attention mechanisms, our model achieves importance assessment and adaptive fusion of modal features, while the designed dimensional compression mechanism significantly reduces computing and memory overhead while maintaining representation capabilities. As a result, MambaRec performed better than the existing baseline model on all three datasets.

(2) \textbf{Most benchmark models suffer from noise interference and insufficient extraction of effective information when processing modal features.}
Specifically, BPR \cite{rendle2012bpr} and LightGCN \cite{he2020lightgcn} completely ignore multimodal content, which makes it difficult for them to cope with cold starts and sparse content-based scenarios, and while VBPR \cite{he2016vbpr} introduces visual features, it lacks structural modeling and has limited representation capabilities. Although MMGCN \cite{wei2019mmgcn}, GRCN \cite{wei2020graph}, and other graph neural network methods integrate graph structure and modal features, they use static or linear fusion strategies, which makes it difficult to effectively filter out noise information. Some methods attempt to alleviate the problems faced. SLMRec \cite{tao2022self} enhances modal discrimination through self-supervision tasks, and FREEDOM \cite{zhou2023tale}introduces a graph structure freezing strategy to reduce error propagation. Problems of noise interference and insufficient information extraction still exist. In contrast, our proposed MambaRec model adaptively filters noise and enhances key modal features by introducing a DREAM module that combines multi-scale convolution and dual attention mechanisms. Our method has achieved the best recommendation performance.

(3) \textbf{Modal alignment can indirectly alleviate the problems of noise interference between modes and insufficient information extraction.}
Although MGCN \cite{yu2023multi} introduces multimodal information into collaborative modeling, it is difficult to effectively suppress modal noise and extract fine-grained semantics due to the use of linear projection and static fusion strategies. In contrast, our method introduces local feature enhancement and global distribution alignment mechanisms. The DREAM module enhances critical information and dynamically suppresses redundant features, while MMD loss reduces cross-modal distribution differences and mitigate noise interference at the representation level. The two work together to demonstrate stronger anti-interference capabilities and optimal recommendation performance.

\begin{figure}[t]
  \centering
  \includegraphics[width=\linewidth]{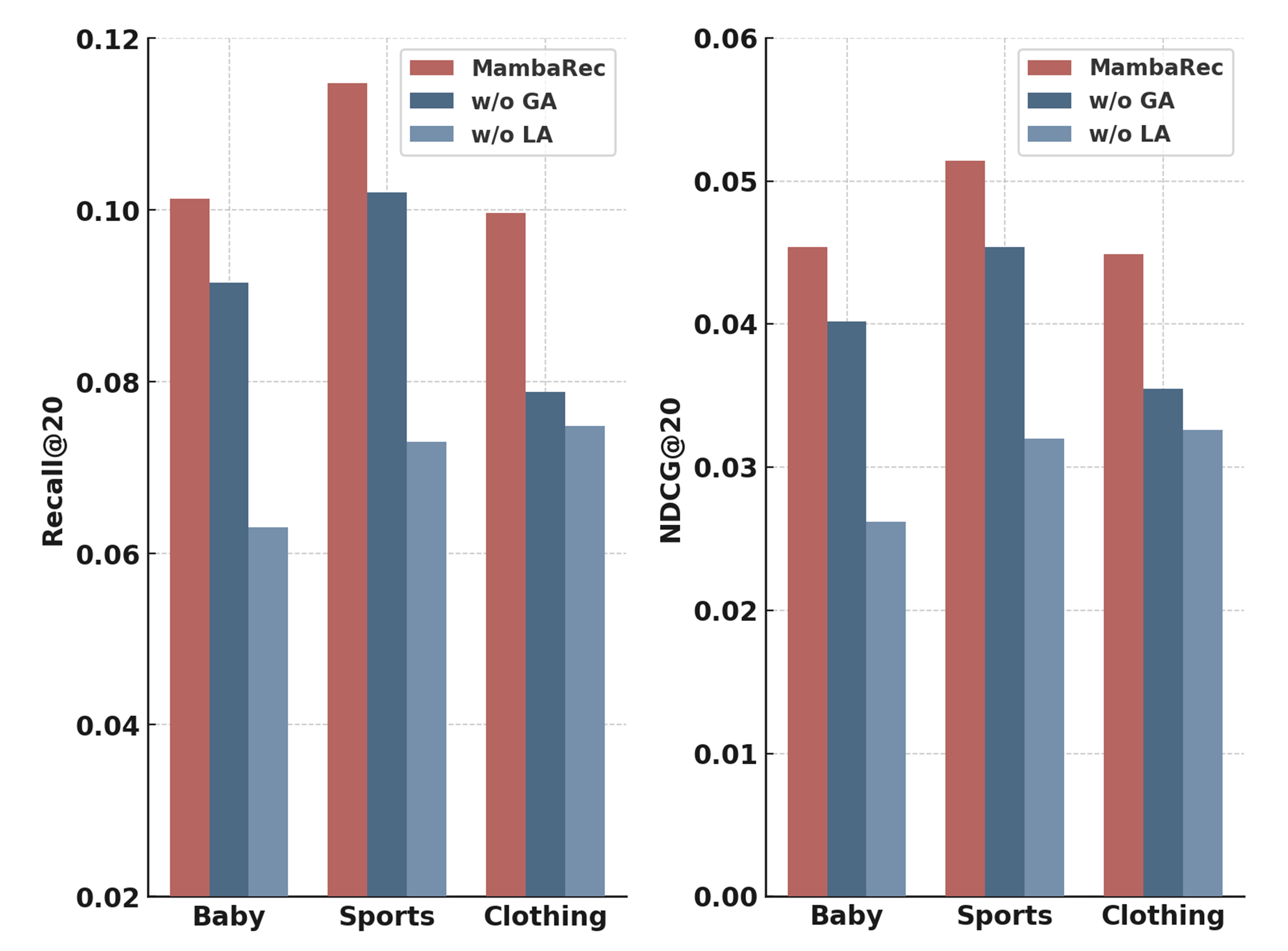}
  \caption{Ablation studies on the proposed MambaRec}
  \Description{The figure presents two side-by-side bar charts comparing the performance of MambaRec with two ablated variants—one without Global Alignment (w/o GA) and the other without Local Alignment (w/o LA)—in terms of Recall@20 and NDCG@20 across three datasets.}
\end{figure}

\begin{table}[tb]  
  \centering      
  \caption{Performance Comparison on multi‐modalities}
  \label{tab:multi-modal}
  \begin{tabular}{clcccc}
    \toprule
    Datasets & Modality & R@10 & R@20 & N@10 & N@20 \\
    \midrule
    \multirow{3}{*}{Baby} & Text & 0.0561 & 0.0863 & 0.0307 & 0.0384 \\
    & Visual & 0.0485 & 0.0761 & 0.0270 & 0.0341 \\
    & Full & \textbf{0.0660} & \textbf{0.1013} & \textbf{0.0363} & \textbf{0.0454} \\
    \midrule
    \multirow{3}{*}{Sports} & Text & 0.0685 & 0.1024 & 0.0373 & 0.0460 \\
    & Visual & 0.0575 & 0.0852 & 0.0310 & 0.0382 \\
    & Full & \textbf{0.0763} & \textbf{0.1147} & \textbf{0.0416} & \textbf{0.0514} \\
    \midrule
    \multirow{3}{*}{Clothing} & Text & 0.0601 & 0.0900 & 0.0327 & 0.0403 \\
    & Visual & 0.0406 & 0.0627 & 0.0217 & 0.0273 \\
    & Full & \textbf{0.0673} & \textbf{0.0996} & \textbf{0.0367} & \textbf{0.0449} \\
    \bottomrule
  \end{tabular}
\end{table}

\subsection{Ablations Studies (RQ2)}
To verify the effectiveness of each module, we conducted ablation experiments from two dimensions. On the one hand, we analyze the impact of local and global modal alignment modules on recommendation performance, and on the other hand, we explore the contribution of different modal input combinations to model effectiveness.

\subsubsection{Effect of Alignment Modules}
We divide the proposed model into three different variants: (i) MambaRec, (ii) MambaRec without local feature alignment (w/o LA), (iii) MambaRec without global distribution alignment (w/o GA). The results are shown in figure. 3:

The complete MambaRec model performed well on all evaluation metrics and datasets. In contrast, once the modal alignment module is removed, model performance drops significantly. The performance degradation is even more pronounced in the absence of a local modal alignment module, indicating that the module is critical in characterizing modal details and capturing local associations. Although the global alignment module focuses on overall semantic alignment, its absence can also weaken the fusion effect. 

\subsubsection{Effect of Modalities}
In order to further evaluate the impact of each modality, we conducted experiments under different input conditions: text input covers text modal content, visual input contains visual modal related information, and full input combines both text and visual modalities. As shown in Table 3, integrating information from text and visual modalities always leads to optimal performance. In contrast, if the model relies on only a single modality, its recommendation effectiveness will be reduced to varying degrees. This phenomenon reflects the complementary characteristics of multimodal information in describing user preferences. Text modalities can better reflect the semantic attributes of items, while visual modalities intuitively present the appearance characteristics of items, and they have their own advantages in different usage scenarios.

\subsection{Sensitivity Analysis (RQ3)}
\subsubsection{Effects of the weight of $\lambda_{cl}$}
Figure. 4 shows the model's performance on the Baby and Sports datasets, with the comparative learning loss weight $\lambda_{cl}$ ranging from $[10^{-4}, 1]$. It can be observed that when $\lambda_{cl}$ increases from $10^{-4}$ to $0.01$, both Recall@20 and NDCG@20 are steadily improving, and when $\lambda_{cl}$ exceeds $0.01$, model performance drops significantly. The experimental results show that $\lambda_{cl} = 0.01$ is the optimal setting on both datasets. Too small the weight will limit the role of the contrast signal and cannot effectively improve the representation ability. However, too large the weight will cause auxiliary tasks to dominate the training process and weaken the optimization effect of the recommended target. Therefore, moderate contrast loss helps enhance the consistency and discrimination of modal representations, thereby improving the overall recommendation performance of the model.

\begin{figure}[t]
  \centering
  \includegraphics[width=\linewidth]{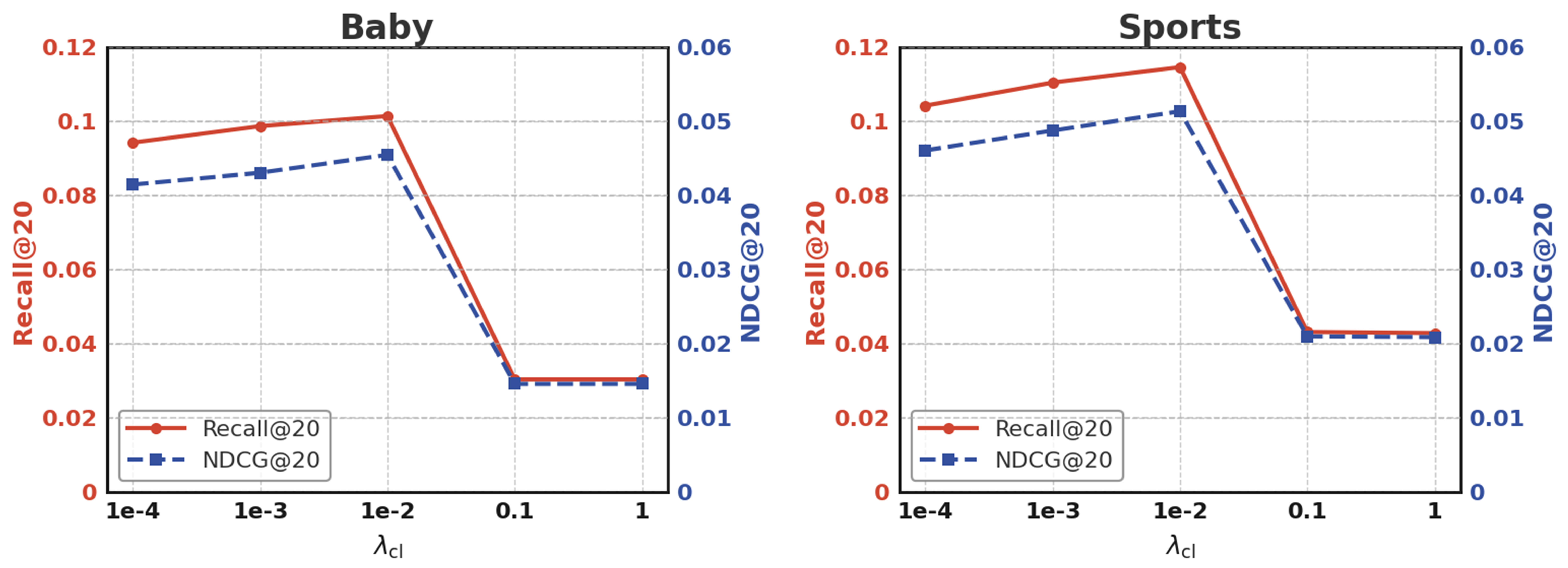}
  \caption{Variation of MambaRec with $\lambda_{cl}$}
  \Description{The figure consists of two rows of line plots, illustrating the impact of the hyperparameter $\lambda_{cl}$ on Recall@20 (red solid line, left y-axis) and NDCG@20 (blue dashed line, right y-axis).}
\end{figure}

\begin{figure}[t]
  \centering
  \includegraphics[width=\linewidth]{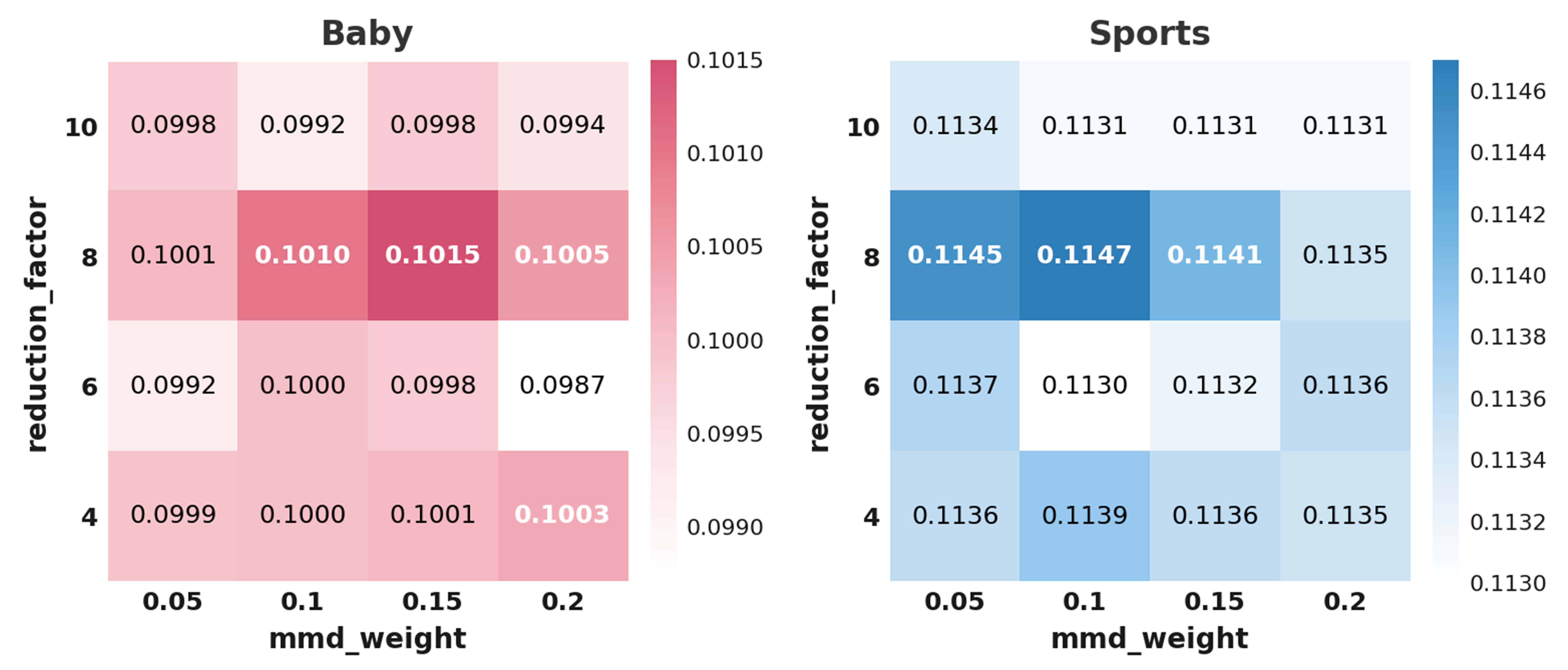}
  \caption{Variation of MambaRec with $\lambda_{mmd}$}
  \Description{The figure contains two side-by-side heatmaps that illustrate the NDCG@20 performance of the model on the Baby (left) and Sports (right) datasets under different combinations of mmd_weight (0.05, 0.1, 0.15, 0.2) and reduction_factor (4, 6, 8, 10).}
\end{figure}

\begin{figure}[b]
  \centering
  \includegraphics[width=\linewidth]{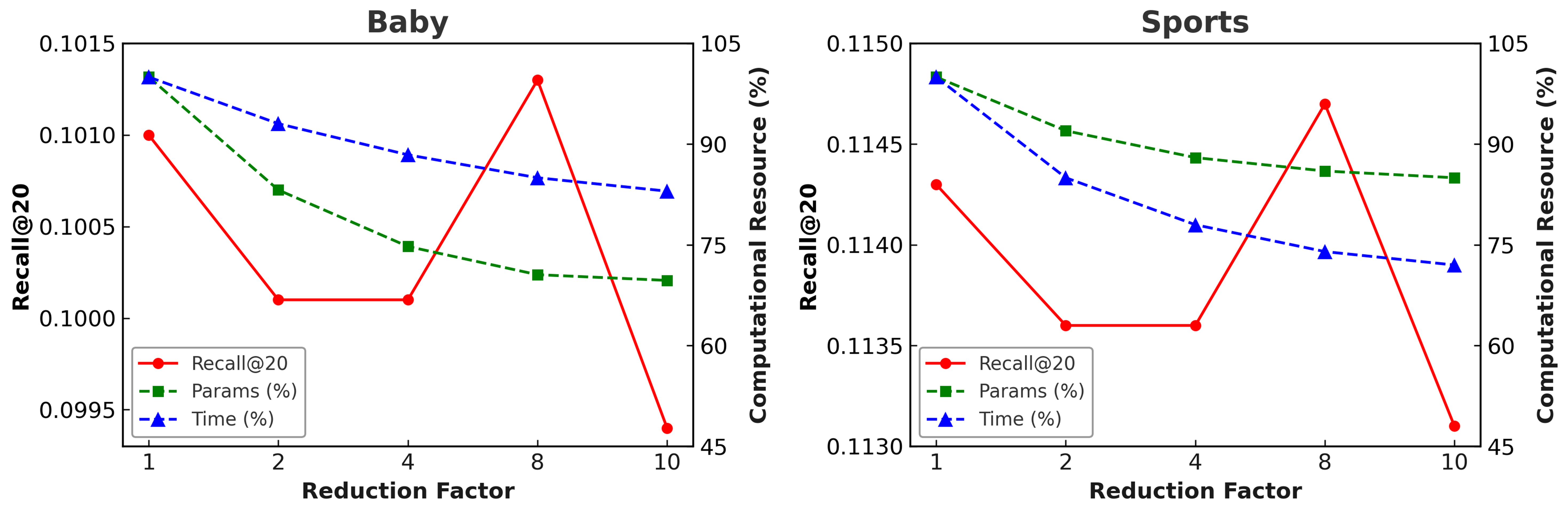}
  \caption{Impact of Reduction Factor}
  \Description{This figure shows the variation of the model's Recall@20 versus the computational resource consumption (number of parameters and inference time) on the two datasets (left: Baby, right: Sports) for different Reduction Factor.}
\end{figure}

\subsubsection{Effects of the weight of $\lambda_{mmd}$}
We compared Recall@20 performance on Baby and Sports datasets under different combinations of reduction factors and maximum mean discrepancy weight ($\lambda_{mmd}$), as shown in Figure. 5. The results show a favorable performance region within the parameter space, with several combinations achieving comparable results. The combination of reduction factor 8 and $\lambda_{mmd}$ = 0.15 performs consistently well on both datasets, though other nearby parameter settings also demonstrate competitive performance. Moderate compression ratios can give better results than excessive compression or almost no compression because moderate compression can remove redundant parameters and noise without serious loss of characterization capabilities. $\lambda_{mmd}$, which is the strength coefficient of modal alignment, also has a significant impact on performance. Too small a value may lead to insufficient alignment of different modal features and ineffective information fusion, while too large a value may introduce irrelevant noise and reduce model discrimination capabilities.

\subsubsection{Effects of Reduction Factor}
To balance model performance and computing resource consumption, we examined three indicators under different compression factors for Baby and Sports datasets, namely, Recall@20, model parameter count as percentage of benchmark, and computation time as percentage of benchmark, as shown in Figure. 6. Experimental results demonstrate that as compression factor increases, model parameters and computation time exhibit continuous decline, while Recall@20 shows nonlinear fluctuation patterns, decreasing at moderate compression levels. Uncompressed models achieve superior performance but incur highest resource costs. Excessive compression significantly reduces parameters and time, but insufficient representation capability leads to notable performance degradation. At compression factor 8, Recall@20 reaches peak levels while reducing parameters by over 30$\%$ and significantly decreasing computation time. Model compression not only reduces resource burden, but also suppresses overfitting noise to a certain extent, thereby improving generalization stability \cite{han2015deep, cheng2017survey}. It is important to choose a reasonable compression factor in resource-constrained deployment scenarios \cite{lian2020lightrec}.

\subsection{Visualization Analysis (RQ4)}
To verify the effectiveness of MambaRec in multimodal fusion, we selected the classic multimodal model VBPR as a baseline for comparison, used t-SNE \cite{van2008visualizing} to reduce the dimensions of the fusion features to two dimensions, and visualized their distribution through Gaussian kernel density estimation, as shown in Figure. 7 and Figure. 8. In these figures, the Baby dataset is represented in red and the Sports dataset in blue. 

From the visualization results, the embedding distribution of the VBPR model shows significant feature clustering effects and multi-modal concentration patterns, which indicates that the model suffers from representation degradation problem \cite{gao2019representation, qiu2022contrastive} when processing multimodal features, where different semantic categories of information concentrate in overlapping regions of the embedding space, making effective discrimination challenging. In contrast, MambaRec has a more uniform embedding distribution, wide coverage and a smooth kernel density estimation (KDE) \cite{terrell1992variable} curve, indicating a more discriminatory representation. In summary, the modal alignment mechanism not only fundamentally alleviates the semantic degradation and modal overlap problems that are easy to occur in traditional methods, but also significantly improves the overall performance of the multimodal recommendation system in terms of expression power and representation structure, which verifies the effectiveness and superiority of the MambaRec model.

\begin{figure}[tb]
  \centering
  \includegraphics[width=\linewidth]{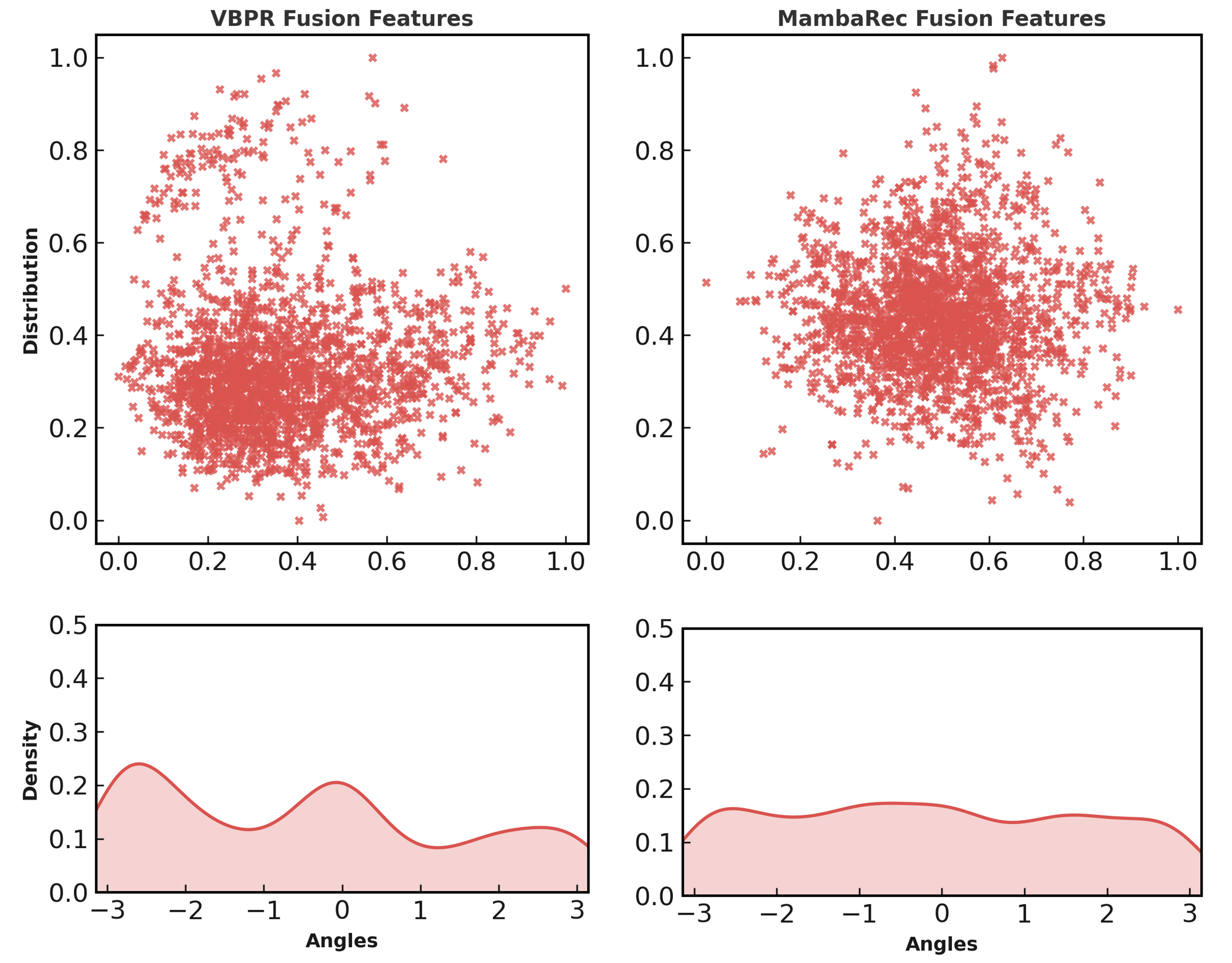}
  \caption{Distribution of representations for Baby Datasets}
  \Description{This figure shows the distribution of Fusion Features generated by the VBPR and MambaRec methods on the Baby dataset.}
\end{figure}

\begin{figure}[tb]
  \centering
  \includegraphics[width=\linewidth]{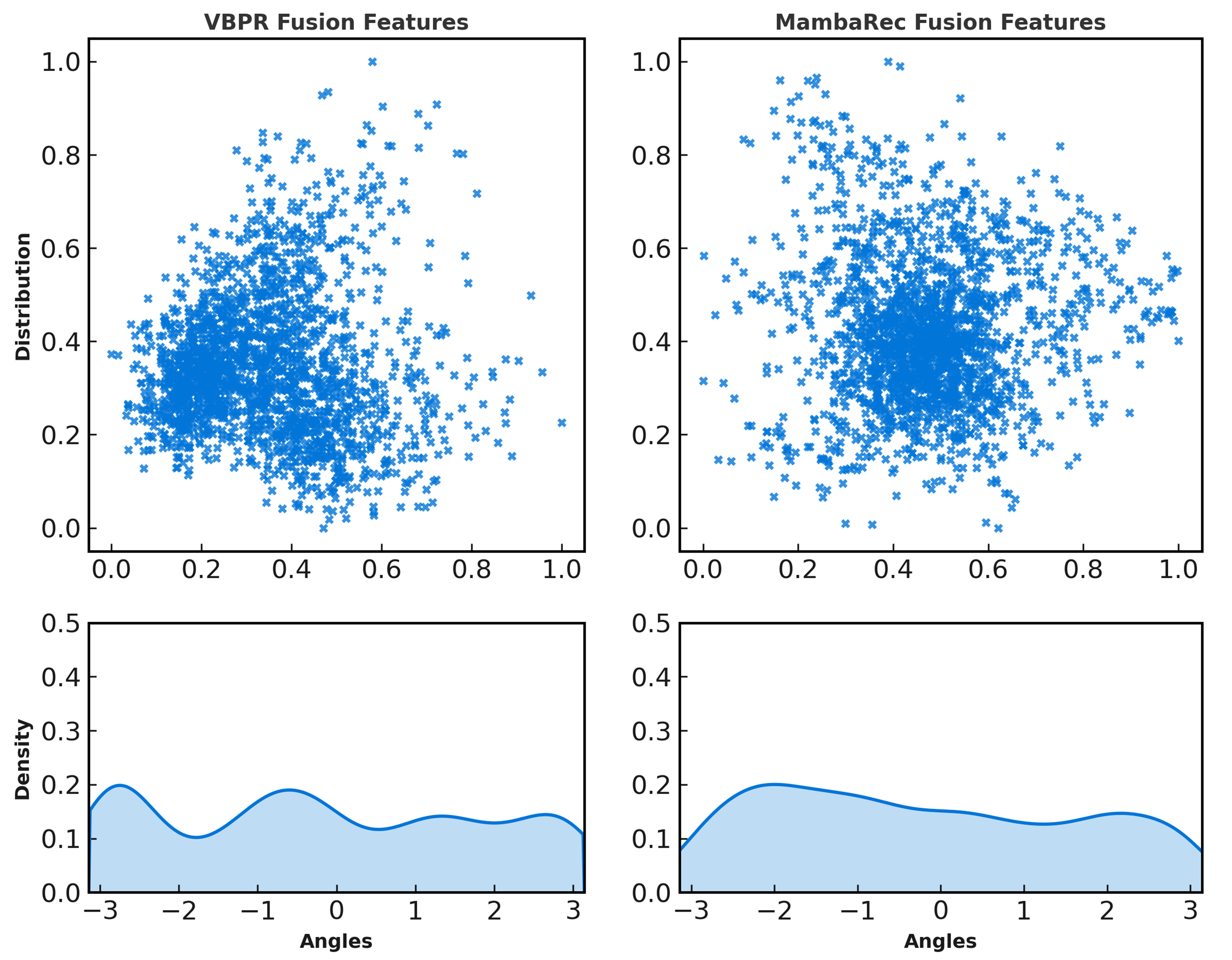}
  \caption{Distribution of representations for Sports Datasets}
  \Description{This figure shows the distribution of Fusion Features generated by the VBPR and MambaRec methods on the Sports dataset.}
\end{figure}

\section{Conclusion}
In this paper, we propose MambaRec, a novel multimodal recommendation framework that achieves local feature alignment via the Dilated Refinement Attention Module (DREAM)—a mechanism that leverages multi-scale dilated convolutions combined with channel-wise and spatial attention to mitigate fusion challenges arising from modality heterogeneity. To ensure global distribution-level alignment, MambaRec integrates Maximum Mean Discrepancy (MMD) loss and contrastive learning to enforce semantic consistency across modalities. Moreover, a tailored dimensionality reduction scheme is introduced to substantially reduce memory consumption and enhance model scalability and deployment efficiency. Overall, MambaRec offers an efficient, robust, and scalable solution for multimodal recommendation, with strong potential for extension to more complex modalities such as video and audio in future work.

\begin{acks}
This work was partly supported by Institute of Information \& communications Technology Planning \& Evaluation (IITP) grant funded by the Korea government (MSIT) (No.RS-2022-00155885, Artificial Intelligence Convergence Innovation Human Resources Development (Hanyang University ERICA)) and the MSIT (Ministry of Science and ICT), Korea, under the Convergence security core talent training business support program (IITP-2024-RS-2024-00423071) supervised by the IITP(Institute of Information \& Communications Technology Planning \& Evaluation) and the MSIT (Ministry of Science, ICT), Korea, under the National Program for Excellence in SW),  supervised by the IITP (Institute of Information \& communications Technology Planing \& Evaluation) in 2025(2024-0-00058).
\end{acks}

\section*{GenAI Usage Disclosure}
This work has been moderately polished using the generative AI tool ChatGPT to improve the clarity and accuracy of the language. All research content, data analysis and academic opinions are independently conceived and completed by us.

\bibliographystyle{ACM-Reference-Format}
\balance
\bibliography{main}










\end{document}